\documentclass[12pt,preprint]{aastex}

\tighten
\newcommand\be{\begin{equation}}
\newcommand\ee{\end{equation}}
\newcommand\jcd{Christensen-Dalsgaard}

\usepackage{epsfig}
\begin{document}
\shortauthors{Antia \& Basu}
\shorttitle{Solar tachocline properties}

\title{Revisiting the solar tachocline: Average properties and temporal variations}
\author{H. M. Antia}
\affil{Tata Institute of Fundamental Research,
Homi Bhabha Road, Mumbai 400005, India}
\email{antia@tifr.res.in}
\and
\author{Sarbani Basu}
\affil{Department of Astronomy, Yale University, P. O. Box 208101,
New Haven CT 06520-8101, U.S.A.}
\email{sarbani.basu@yale.edu}

\begin{abstract}
The tachocline is believed to be the region where the solar dynamo operates.
With over a solar cycle's worth of data available from the MDI and GONG 
instruments, we are in a position to investigate not merely the average
structure of the solar tachocline, but also its time variations.
We determine the properties of the tachocline as a function of time
by fitting a two-dimensional model that takes latitudinal variations
of the tachocline properties into account.
We confirm that if we consider central position of the tachocline,
it is prolate. Our results show that the tachocline 
is thicker at higher latitudes than the equator, making the overall 
shape of the tachocline more complex.  Of the tachocline
properties examined, the transition of the rotation rate across
the tachocline, and to some extent the position of the tachocline,
show some temporal variations. 
\end{abstract}


\section{Introduction}
\label{sec:intro}

Inversions of helioseismic data have shown that the solar convection zone 
rotates differentially, but that the radiative interior has an almost
solid-body like rotation (see Schou et al.~1998 and references therein).
The transition from differential rotation to solid body rotation occurs
over a very narrow region, that is called the tachocline (Spiegel \& Zahn 1992).
The tachocline is believed to be the seat of the solar dynamo, and hence
is expected to affect, and be affected by, the solar dynamo.

The poor resolution of the inversion techniques,
particularly at high latitudes, makes it
difficult to infer the properties of the tachocline from normal
inversions for the solar rotation rate.
One consequence of the poor resolution is that the transition
appears to occur over a large radial distance.
 As a result, tachocline studies usually involve forward modelling techniques,
with the parameters of the models determined by fitting helioseismic data
(e.g., Kosovichev 1996; Basu 1997; Antia et al.~1998; Charbonneau et al.~1999;
Basu \& Antia 2001, 2003). Early investigations have shown that the
tachocline is prolate. Also clear (even from inversions) is that
the `jump' in the rotation rate across the tachocline is latitude dependent.
At the equator, the rotation rate changes from a higher value to a lower value
when moving from the convection zone to the radiative interior through the tachocline,
there is almost no change around a latitude of about $30^\circ$, while
at higher latitudes the rotation rates shifts from a lower value to
a higher one.  What is not completely clear from  earlier work however, is whether the
tachocline thickness, i.e., the radial distance over which the rotation rate
changes, is larger at higher latitudes than the equator.
Although, the early results found that the thickness does increase with
latitude, the statistical significance of this increase was not clear.

The bulk of the solar convection zone shows clear, periodic, changes in the rotation
rate in the form of zonal flows (Vorontsov et al.~2002; Basu \& Antia 2003;
Howe et al.~2005). It is, however, not clear if the
tachocline too changes with time. Early studies by Basu \& Antia~(2001; 2003)
using data available for a limited part of the solar cycle 23 during its
ascending phase did not find any significant temporal variation. However, 
with more than a solar cycle worth of data available now, we revisit the
question of tachocline variations. We also examine whether the tachocline structure
was different between the solar minimum preceding cycle 23 and the exceptionally
long and deep minimum that preceded cycle 24.
We also use the increased amount of data  to 
determine the average properties of the tachocline more accurately.

\section{Data and analysis}
\label{sec:data}

We use data obtained by the GONG (Hill et al.~1996) and MDI
(Schou 1999) projects for this work.
These data sets consist of the mean frequency and the splitting coefficients
of different $(n,\ell)$ multiplets.
Only the odd-order splitting coefficients are needed to determine
the rotation rate in the solar interior (e.g., Ritzwoller \& Lavely 1991).
We use 151 data sets from GONG, each set
covering a period of 108 days. The first set starts on
1995, May 7 and the last set ends on 2010, June 4, with a
spacing of 36 days between consecutive data sets.
Thus these sets cover about a year of data leading to the minimum of
cycle 23 as well as about a year of data following the minimum of cycle 24.
The MDI data  consist of 73 non-overlapping
sets each obtained from observations taken over a   period of
72 days. The first set begins on 1996, May 1 and the last
set ends on  2011, February 11. The MDI data start close to the minimum
of cycle 23 and do not cover preceding period, while these data cover
a period of more than a year following the minimum of cycle 24.

To determine the properties of the tachocline we use the 2D-annealing
technique described by Antia et al.~(1998). In this method we fit a
rotation rate of the form
\begin{equation}
\Omega_{\rm tac}(r,\theta)=\cases{\Omega_c
+\frac{\delta\Omega}{1+\exp[(r_t-r)/w]}& if $r\le0.70$\cr
\Omega_c+B(r-0.7)
+\frac{\delta\Omega}{1+\exp[(r_t-r)/w]}& if $0.70<r\le0.95$\cr
\Omega_c+0.25B-C(r-0.95)
+\frac{\delta\Omega}{1+\exp[(r_t-r)/w]}& if $r>0.95$\cr}
\label{eq:tach}
\end{equation}
where $r$ is the radial distance in units of solar radius, $\theta$ is the co-latitude
and
\begin{eqnarray}
B&=&B_1+B_3P_3(\theta)+B_5P_5(\theta),\\
\delta\Omega&=&\delta\Omega_1+\delta\Omega_3P_3(\theta)+\delta\Omega_5P_5(\theta),\\
r_t&=&r_{d1}+r_{d3}P_3(\theta),\\
w&=&w_1+w_3P_3(\theta),\\
P_3(\theta)&=&5\cos^2\theta-1,\\
P_5(\theta)&=&21\cos^4\theta-14\cos^2\theta+1.
\end{eqnarray}
The quantity $r_t$ is the central point of the transition region in the rotation rate and
what we consider to be the `position' of the tachocline. The half-width of the tachocline
is given by the quantity $w$, and $\delta\Omega$ is the jump in the rotation
rate across the tachocline. The quantities $r_t$, $w$ and $\delta\Omega$
define the tachocline. The form for the smooth part of rotation rate used
here is slightly different from that used by Basu \& Antia (2001).
The difference is in the radiative interior ($r<0.7R_\odot$) where we
assume uniform rotation.
The rotation rate given by Eq.~(\ref{eq:tach}) is used to calculate the
splitting coefficients and these are compared with observed splitting
coefficients.
We find that the first component in $\delta\Omega$ is generally small
and that the fits become more stable if that is ignored,  hence all results in
this work are obtained by setting $\delta\Omega_1=0$.
The parameters $r_{d1}$,$r_{d3}$,$w_1$,$w_3$,$\delta\Omega_3$,
$\delta\Omega_5$,$B_1$,$B_3$,$B_5$,$C$, and $\Omega_c$ are fitted to the observed splitting
coefficients. Only the first 3 odd-order splitting coefficients are used for
this purpose. We use the method of simulated annealing to perform the
least squares fit to the observed frequencies.
Note that our model of the tachocline differs
from that of Kosovichev (1996) and Charbonneau et al.~(1999) and to compare
the width we obtain to those obtained by them, $w$ needs to be multiplied 
by a factor of $2.5$.

For each available data set in GONG and MDI data sets we fit rotation rate
of the form given by Eq.~(\ref{eq:tach}) to the observed splitting coefficients
to calculate the 11 parameters of the model.
In this work we are only interested in the parameters that define the properties of
the tachocline.

\section{Results and discussion}
\label{sec:results}

\subsection{The time-averaged tachocline}

The time averaged parameters for the tachocline are listed in Table~1.
The average tachocline properties were determined by averaging all
results.
GONG and MDI results were averaged, and are listed,  separately.
The average results are consistent with those of Basu \& Antia (2001) obtained using a
subset of current data and with a slightly different form for the rotation
rate.
As can be seen, that there are some systematic differences between the
GONG and MDI results. Similar
differences have been seen in rotation inversion results (Schou et al.~2002).
The discrepancy is mainly attributed to differences in the data processing
pipelines of the two projects. Studies of temporal variations in solar
rotation rate (e.g., Antia et al.~2008a) suggest that the time-varying
component of the rotation rate is not affected by the discrepancy
between GONG and MDI data.

The fitted parameters can be used to calculate the properties of the
tachocline at any given latitude. The time-averaged results at a few
latitudes are shown in Table~2. As can be seen clearly, the 
tachocline is prolate as measured
by the central radius defining the tachocline. GONG data yields
a difference of $(0.012\pm0.002)R_\odot$ between the latitude of $60^\circ$ and the equator.
The difference with MDI data is $(0.040\pm0.003)R_\odot$.
Our results are in agreement with earlier
results (Antia et al.~1998; Corbard et al.~1998, 1999; Charbonneau et al.~1999;
Basu \& Antia 2001, 2003). 

Table~2 also shows a clear variation of the thickness of the tachocline.
What we find is that the tachocline is the least thick at the equator, and the
thickness increases steadily with latitude. While earlier investigations
(Charbonneau et al.~1999; Basu \& Antia 2001, 2003) had indicated that this
might be the case, the results were not statistically significant. The larger
quantity of data available now has resolved this issue. 
We find that the thickness increases by about  $(0.018\pm0.003)R_\odot$ between the
equator and $60^\circ$ latitude as per GONG and $(0.028\pm0.004)R_\odot$
as per MDI data.

While the central part of the tachocline (as define by $r_t$) is clearly prolate in shape, 
the overall shape of the tachocline is more complex since the thickness changes. 
Given the model of the tachocline (Eq.~\ref{eq:tach}), we can assume that the
tachocline is bounded between layers with radius $r_t-2w$ and $r_t+2w$.
The rotation rate changed by about 76\% of  $\delta\Omega$
within these limits.  The upper boundary of the tachocline is
clearly prolate since both $r_t$ and $w$ increase with latitude. The
lower boundary is another matter. Given the tachocline model adopted in this work,
the lower boundary is determined by the value
of $r_{t3}-2w_3$, which is $-(0.0071\pm0.0021)R_\odot$ (GONG) or
$-(0.0066\pm0.0030)R_\odot$ (MDI). Thus within  2--3$\sigma$,
the lower boundary is close to being spherical. A similar
conclusion was reached by Basu \& Antia (2003). 
It should be noted that the
base of the convection zone is also spherical (Basu \& Antia 2001).
The lower boundary of the tachocline is at about $0.68R_\odot$ which is
consistent with the extent of mixing required below the solar convection zone 
to match the solar sound-speed profile (e.g., Brun et al.~2002).

The latitudinal variation in $\delta\Omega$
is very clear. There is very little change in
the rotation rate across the tachocline at a latitude of about $30^\circ$.
At lower latitudes $\delta\Omega$ is positive (i.e., higher rotation rate above the tachocline),
while at higher latitudes the sign of the difference is reversed.
The latitude at which $\delta\Omega$ changes sign is of some interest.
With the parameters of tachocline as determined by us this turns out
to be about $29^\circ$. If $P_3(\theta)$ were the 
only term present in the definition of $\delta\Omega$, we
would expect this number to be $26.6^\circ$. Given that $P_3(\theta)$
is the dominant term, it is not surprising that the latitude at which
$\delta\Omega=0$ is close to this value. The latitude at which $\delta\Omega=0$ can
also be determined quite easily by inspecting solar rotation profiles
obtained from inversions. The numbers are quite similar.

\subsection{Activity and time dependence}

In order to detect tachocline variations linked to solar activity, we also average
results that correspond to times of high and low activity --- we define the
period of high activity to be the one for which the 10.7 cm radio flux was greater than 140 SFU
and the period of low activity is defined as the ones with 10.7 cm flux
less than 90 SFU.
To study possible difference between the two periods of minimum activity
covered by the data sets we also take separate averages for the two periods
of low activity. These are also listed in Table~1.

We do not find any significant change in either the
thickness or the position  of the tachocline between the high and low activity periods.
However,
$\delta\Omega_3$ and $\delta\Omega_5$ show differences at the level
of 2--3$\sigma$. To check for any solar cycle variation we also calculate
the correlation coefficient between these parameters and the 10.7 cm radio
flux. The correlation coefficients for 
$\delta\Omega_3$ and $\delta\Omega_5$ are found to be 0.40 and 0.20
respectively, when GONG data is used. For MDI data these correlations are
0.36 and 0.40 respectively. The correlation with other parameters of the
tachocline are found to be very small. 

GONG and MDI give somewhat disparate results about the differences in
the tachocline between the minimum before cycle 23 and that before cycle 24. 
This could be a result of the fact that MDI data do not completely
span the cycle 23 minimum. 
The only significant change is in the
parameter $\delta\Omega_5$. Even that is of opposite sign between GONG and
MDI. Thus we cannot say for certain whether the
tachocline parameters were different between the two minima.

We take a closer look at the changes in tachocline parameters by
examining a few latitudes in more detail. For each latitude we
have determined the correlation coefficient between $r_t$, $\delta\Omega$
and $w$ and the 10.7 cm radio flux. However, since the 10.7 cm flux is 
an indicator of global activity, while the changes in the tachocline
may be correlated with other local changes, which depend on latitude,
we also
fit an oscillatory form with a period of solar cycle. Thus
for $\delta\Omega$ we fit
\begin{equation}
\delta\Omega(t,\theta)=\langle\delta\Omega\rangle +\sum_{k=1}^3
a_k(\theta)\sin(k\omega_0 t+\phi_k)
\label{eq:osc}
\end{equation}
where $\omega_0$ is the frequency corresponding to 11.7 years, the dynamical length
of the solar cycle as determined by
Antia \& Basu~(2010), and the angular brackets denote
average over time. 
This form was motivated by the fact that the change of zonal
flow velocities at a given latitude and radius can be fitted with this expression.
We fit similar expressions to $r_t$ and $w$.

\begin{figure}
\epsscale{0.9}
\plotone{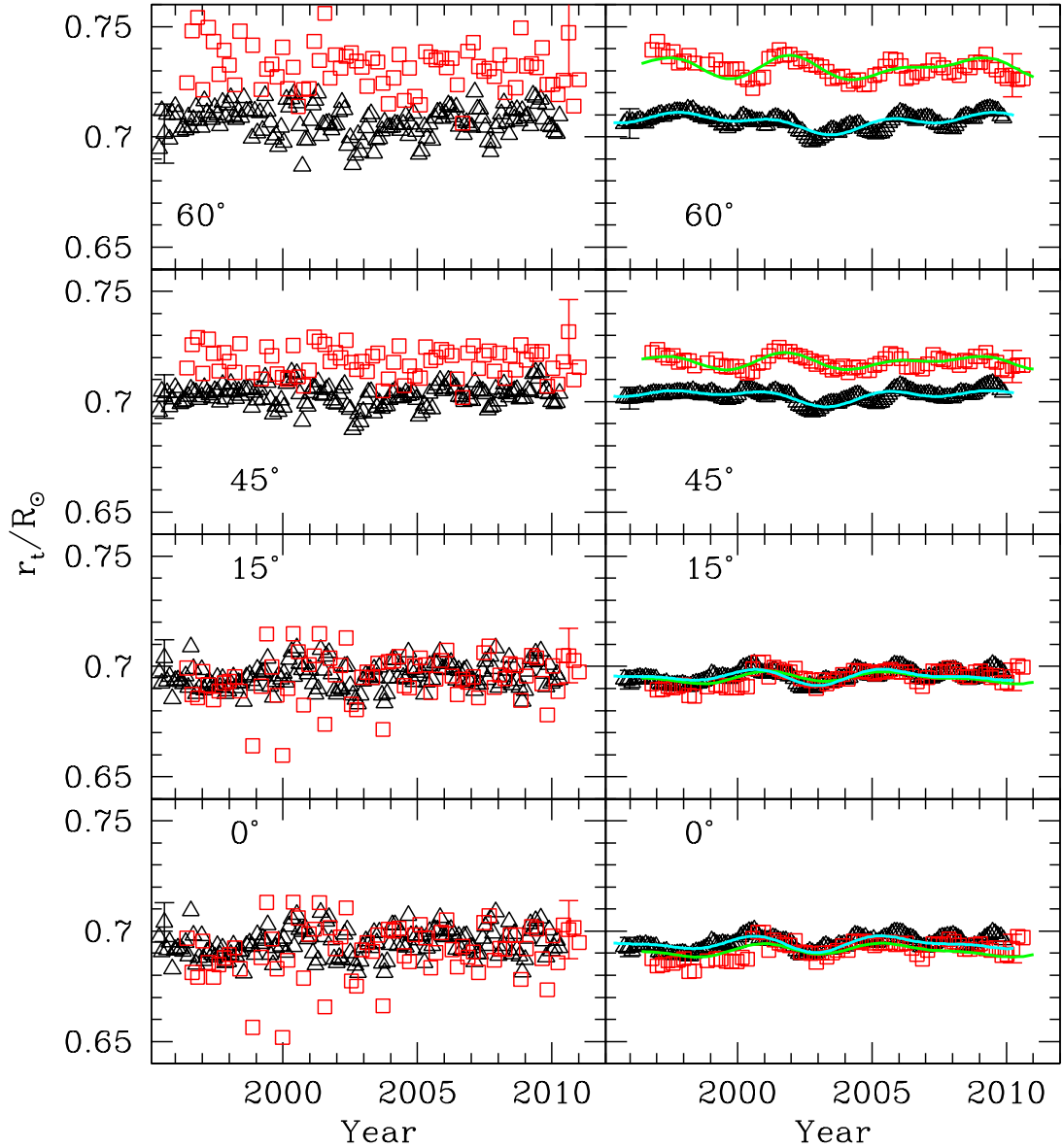}
\caption{The mean position of the tachocline at a few different latitudes
plotted as a function of time. The left panels show
the results for each data set, while the right panel shows running mean
over about a year of data, as well as fits to oscillatory form. The red points are
results with MDI data and black are those with GONG data. We show only 
one representative error-bar each for the GONG and MDI sets for the
sake of clarity.}
\label{fig:rd}
\end{figure}

\begin{figure}
\epsscale{0.9}
\plotone{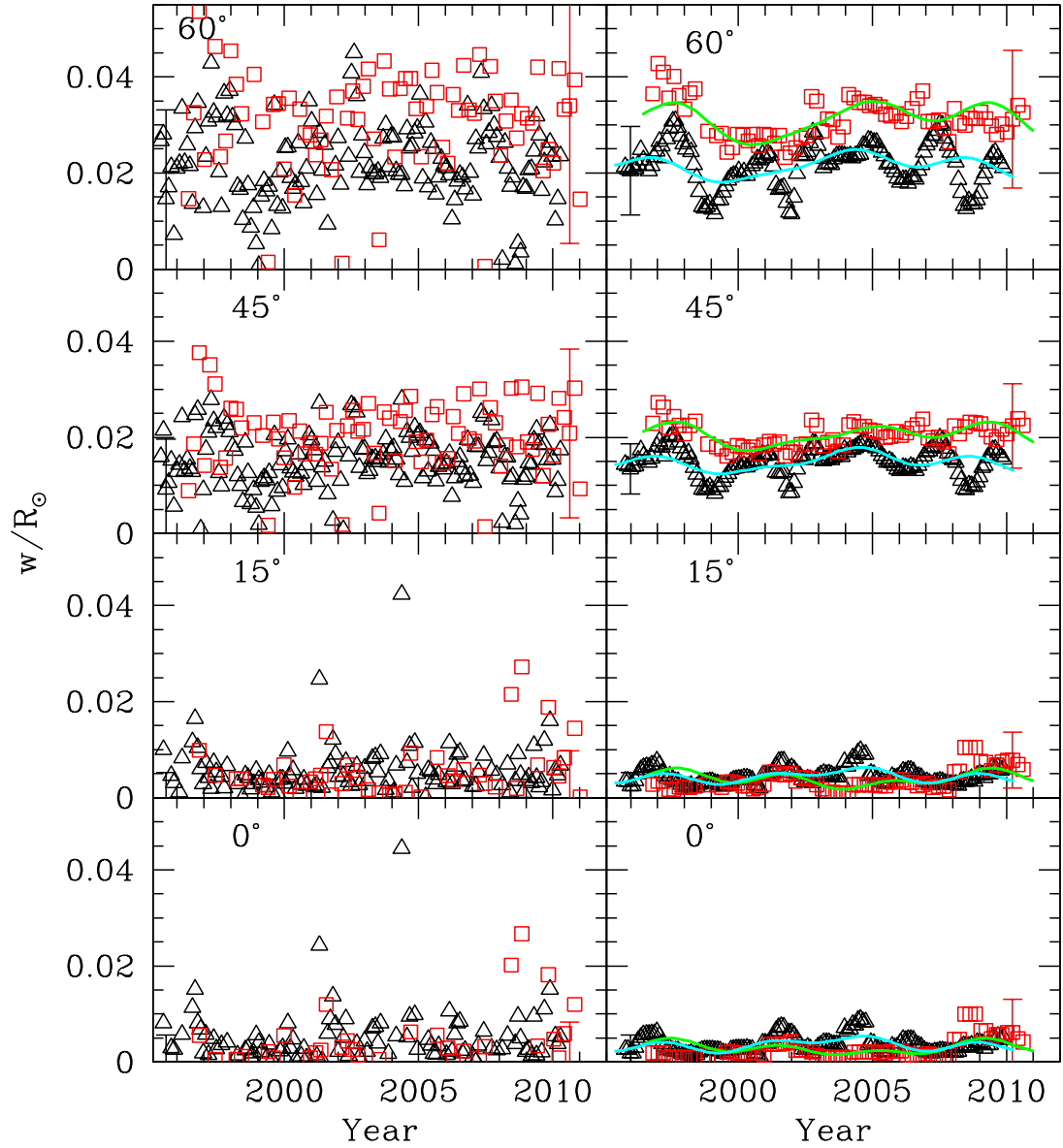}
\caption{Similar to Fig.~\ref{fig:rd}, but showing $w$, the half-width of the tachocline,
instead.}
\label{fig:wid}
\end{figure}

In Figure~\ref{fig:rd} we show the position of the tachocline, $r_t$ at four latitudes
plotted as a function of time. As can be seen, there is a large spread in the results
and there is no clear temporal variation. In the same figure we
also show a running mean of the results to reduce the scatter. The running mean is taken over a period of about
1 year, i.e., over 9 sets for GONG and 5 sets for MDI. Also
shown is the fit to the form given by Eq.~(\ref{eq:osc}).
There is
a good agreement between the results obtained using GONG and MDI data
at low latitudes, but there are significant differences at high latitudes
as also seen in the time-averaged data.
The $r_t$ values are not correlated with the 10.7 cm flux, the correlation coefficients
lie between $-0.14$ and $0.02$ for the different latitudes and sets of data.
There appears to be an oscillatory time variation at the lower latitudes,
and the results are consistent for both data sets, implying the results
may have some significance.
Figure~\ref{fig:wid} shows similar results for the width. At low latitudes the
width is comparable to errors in individual data points and it is
difficult to say much about the temporal variations. At high latitudes
also the temporal variations are unclear --- there is no agreement between
results obtained with the GONG and MDI data. As with $r_t$, the correlation
with the 10.7 cm flux is small.

\begin{figure}
\epsscale{0.9}
\plotone{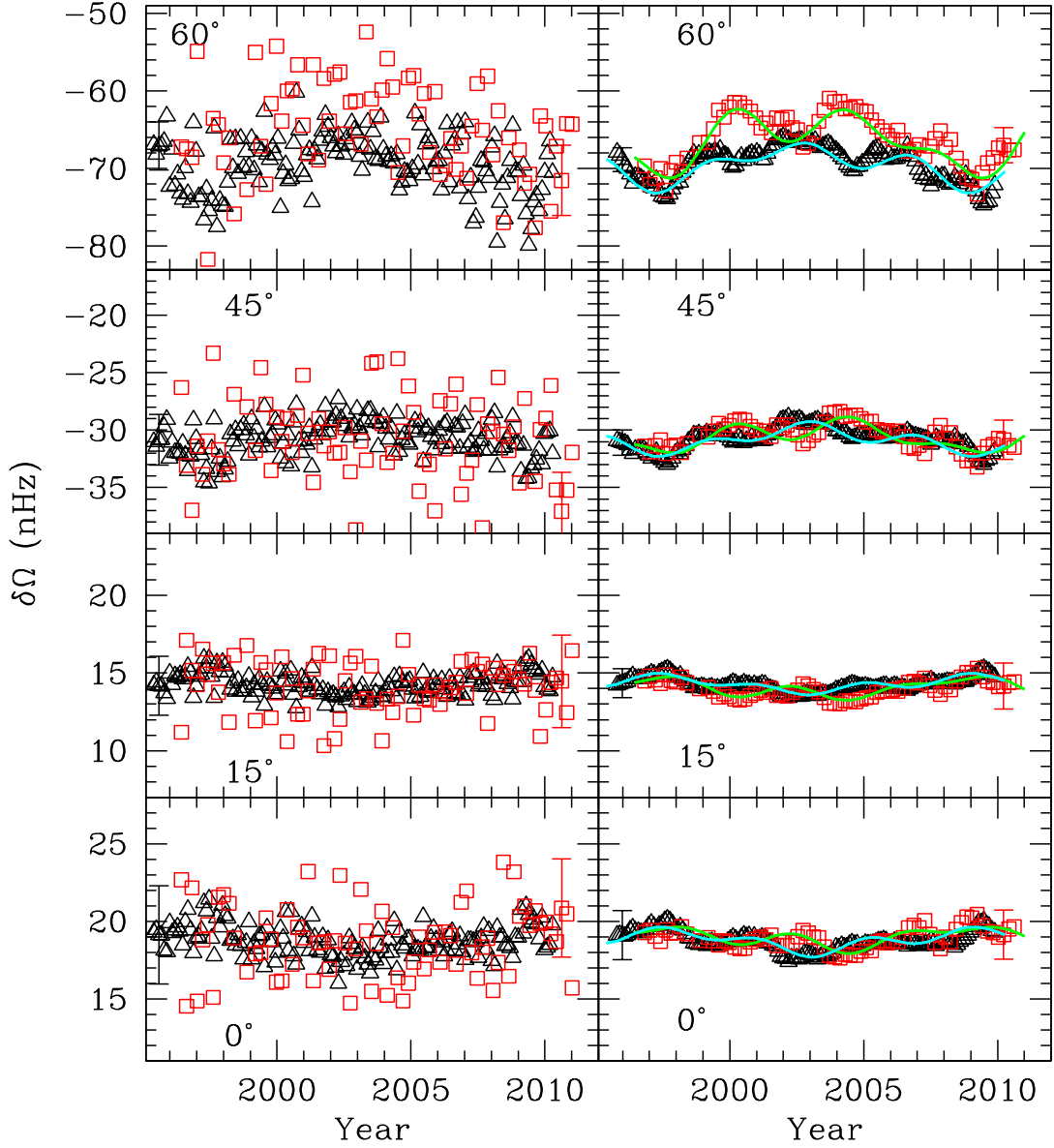}
\caption{Similar to Figs.~\ref{fig:rd}\ and ~\ref{fig:wid}, but showing
$\delta\Omega$, the change in the rotation rate across the tachocline.}
\label{fig:jump}
\end{figure}

Figure~\ref{fig:jump} shows results for $\delta\Omega$. GONG and MDI results agree
at most latitudes, though the agreement is poor at $60^\circ$. This is the only quantity that
shows a reasonable correlation with the 10.7 cm flux, being correlated
with activity at high latitudes and anti-correlated at low latitudes.
The correlation coefficient for latitudes
$0^\circ$, $15^\circ$, $45^\circ$, and $60^\circ$ is
respectively $-0.31$, $-0.39$, $0.36$ and $0.39$ for
the GONG data, and $-0.22$, $-0.34$, $0.20$ and $0.41$ for
MDI. These correlations are consistent with that seen in rotational kinetic
energy in the lower part of the convection zone (Antia et al.~2008b) which
also shows correlation at high latitude and anti-correlation at low latitudes.
Each latitude shows some type of oscillatory behavior in addition
to the solar cycle variation, however,
the oscillatory behavior of the GONG and MDI results do not
agree at any latitude and are out of phase with each other.

Inversion results for rotation rate suggest that the zonal flow pattern penetrates
through the convection zone (Vorontsov et al.~2002; Basu \& Antia 2003;
Howe et al.~2005; Antia et al.~2008a), which could imply some temporal
variations in the tachocline region. However, this variation could  be
in what we consider the smooth part of the rotation rate
(parameters $\Omega_c$, $B$ and $C$ in Eq.~\ref{eq:tach}) and may not 
necessarily affect the tachocline properties $r_t$, $w$ or $\delta\Omega$. 
The temporal variation that we find in $\delta\Omega$ is indeed of the same 
order as the temporal variation in the rotation
rate and it could account for a part of the zonal flow pattern. In this
work we have adopted a crude representation of smooth part of rotation
rate which may not be able to fully represent all the rotation rate variations.
We however, do find that if we compute the zonal flow pattern at a region just
above the tachocline (around $r=0.75R_\odot$) the pattern qualitatively
matches the zonal flow pattern at a comparable depth, though the errors
in both results are too large to see the pattern clearly. 
As a result, we do not show those results.
In addition to the zonal flow pattern, Howe et al.~(2000) found an
oscillatory pattern with a period of 1.3 yrs in the tachocline region.
We do not see any oscillations with a comparable period in the tachocline
properties.
It should be noted however, that we have not been able to
confirm the 1.3 year oscillation (Basu \& Antia 2001, 2003; Antia \& Basu 2010)
even in the zonal flow pattern.

Although we have not examined possible temporal variations in the shape
of the convection zone in this work, given that the errors in
position of the solar convection-zone base are much smaller,
we do not expect the results of Basu \& Antia~(2001) to
be modified by additional data.

\section{Conclusions}
\label{sec:disc}

We have used helioseismic data spanning cycle 23 and beyond to
study the properties of the solar tachocline. We confirm that the center
of the tachocline is prolate in shape. The results show unequivocally that the
thickness of the tachocline increases with increasing latitude, making the
overall shape of the latitude more complex --- while the
outer boundary is prolate, the inner boundary is close to spherical or perhaps
even a little oblate.
This appears to be consistent with the fact that the base of the convection
zone is almost spherical.

The jump across the tachocline is the only parameter that shows a significant
change with solar activity. Other parameters also appear to show some change with time, however
the tachocline properties obtained with GONG data and those with MDI do not 
always show consistent behavior.
If the oscillatory temporal variations of tachocline properties
are confirmed, it would imply that the solar zonal flow (which is the
temporally varying component of solar rotation) penetrates to the base of
the convection zone.

\acknowledgements

This work  utilizes data obtained by the Global Oscillation
Network Group (GONG) project, managed by the National Solar Observatory,
which is
operated by AURA, Inc. under a cooperative agreement with the
National Science Foundation. The data were acquired by instruments
operated by the Big Bear Solar Observatory, High Altitude Observatory,
Learmonth Solar Observatory, Udaipur Solar Observatory, Instituto de
Astrofisico de Canarias, and Cerro Tololo Inter-American Observatory.
This work also utilizes data from the Solar Oscillations
Investigation/ Michelson Doppler Imager (SOI/MDI) on the Solar
and Heliospheric Observatory (SOHO).  SOHO is a project of
international cooperation between ESA and NASA.
SB acknowledges support from NSF grant ATM 0348837
and NASA grant NXX10AE60G.

\clearpage

\begin{table}
\caption{Average properties of the tachocline}
\medskip
{\scriptsize
\begin{tabular}{lcccccc}
\tableline\tableline
Data sets& $r_{t1}/R_\odot$ & $r_{t3}/R_\odot$ & $w_1/R_\odot$ & $w_3/R_\odot$
 & $\delta\Omega_3$ & $\delta\Omega_5$ \\
\tableline
\multicolumn7c{GONG data sets}\\
All sets&         $0.6978\pm0.0011$&$0.0033\pm0.0006$& $0.0071\pm0.0005$&$0.0052\pm0.0010$& $-22.28\pm0.12$&$-3.52\pm0.06$\\
$F_{10.7}>140$ sfu&$0.6974\pm0.0020$&$0.0029\pm0.0010$& $0.0068\pm0.0009$&$0.0048\pm0.0017$& $-21.76\pm0.21$&$-3.34\pm0.11$\\
$F_{10.7}<90$ sfu& $0.6985\pm0.0016$&$0.0036\pm0.0008$& $0.0072\pm0.0007$&$0.0052\pm0.0013$& $-22.66\pm0.17$&$-3.61\pm0.09$\\
cycle 23 min&     $0.6965\pm0.0026$&$0.0040\pm0.0014$& $0.0076\pm0.0012$&$0.0055\pm0.0022$& $-22.75\pm0.28$&$-3.37\pm0.15$\\
cycle 24 min&     $0.6994\pm0.0020$&$0.0033\pm0.0010$& $0.0070\pm0.0009$&$0.0051\pm0.0017$& $-22.62\pm0.21$&$-3.74\pm0.11$\\
\tableline
\multicolumn7c{MDI data sets}\\
All sets&         $0.7021\pm0.0015$&$0.0106\pm0.0007$& $0.0076\pm0.0012$&$0.0086\pm0.0013$& $-21.72\pm0.15$&$-2.77\pm0.08$\\
$F_{10.7}>140$ sfu&$0.7029\pm0.0024$&$0.0106\pm0.0012$& $0.0070\pm0.0020$&$0.0072\pm0.0020$& $-21.20\pm0.24$&$-2.42\pm0.12$\\
$F_{10.7}<90$ sfu& $0.7026\pm0.0020$&$0.0106\pm0.0010$& $0.0086\pm0.0016$&$0.0086\pm0.0017$& $-22.34\pm0.20$&$-3.01\pm0.10$\\
cycle 23 min&     $0.7002\pm0.0040$&$0.0137\pm0.0019$& $0.0063\pm0.0032$&$0.0112\pm0.0033$& $-22.67\pm0.40$&$-3.46\pm0.19$\\
cycle 24 min&     $0.7034\pm0.0024$&$0.0098\pm0.0012$& $0.0093\pm0.0020$&$0.0079\pm0.0020$& $-22.24\pm0.24$&$-2.89\pm0.12$\\
\tableline
\end{tabular}
}
\end{table}

\begin{table}
\caption{Latitudinal variation in the tachocline properties}
{\footnotesize
\vspace{0.5 true cm}
\begin{tabular}{lccrccr}
\tableline\tableline
Lat&\multicolumn3c{GONG data}&\multicolumn3c{MDI data}\\
&$r_{t}/R_\odot$ & $w/R_\odot$ & $\delta\Omega$ (nHz) &
$r_{t}/R_\odot$ & $w/R_\odot$ & $\delta\Omega$ (nHz) \\
\tableline
$0^\circ$ &$0.6945\pm0.0013$&$0.0037\pm0.0011$&$ 18.76\pm0.45$&   $0.6915\pm0.0016$&$0.0028\pm0.0017$&$18.95 \pm0.39$\\
$15^\circ$&$0.6956\pm0.0012$&$0.0045\pm0.0008$&$ 14.27\pm0.27$&   $0.6951\pm0.0016$&$0.0038\pm0.0015$&$14.01 \pm0.37$\\
$30^\circ$&$0.6986\pm0.0012$&$0.0084\pm0.0006$&$-1.39\pm 0.27$&   $0.7048\pm0.0015$&$0.0097\pm0.0013$&$-2.14 \pm0.37$\\
$45^\circ$&$0.7027\pm0.0014$&$0.0149\pm0.0015$&$-30.78\pm0.32$&   $0.7181\pm0.0018$&$0.0204\pm0.0022$&$-30.50\pm0.43$\\
$60^\circ$&$0.7068\pm0.0017$&$0.0215\pm0.0027$&$-69.40\pm0.44$&   $0.7313\pm0.0025$&$0.0311\pm0.0036$&$-66.12\pm0.57$\\
$75^\circ$&$0.7098\pm0.0024$&$0.0264\pm0.0035$&$-103.53\pm0.64$&  $0.7410\pm0.0030$&$0.0390\pm0.0047$&$-96.80\pm0.79$\\
\tableline
\end{tabular}
}
\end{table}

\clearpage

\clearpage

\clearpage

\end{document}